\documentclass[aps,pra,twocolumn,superscriptaddress]{revtex4-1}
\usepackage{xcolor, graphicx}
\usepackage{amsmath, amssymb}
\usepackage[colorlinks=true,urlcolor=blue,citecolor=blue,linkcolor=magenta]{hyperref} %%%%% added options
\usepackage{textcomp}
\usepackage{braket,ulem}

\begin{document}

\title{Probing free-space quantum channels with laboratory-based experiments}

\author{M. Bohmann}
\email{martin.bohmann@uni-rostock.de}
\affiliation{Arbeitsgruppe Theoretische Quantenoptik, Institut f\"ur Physik, Universit\"at Rostock, D-18051 Rostock, Germany}

\author{R. Kruse}
\email{regina.kruse@uni-paderborn.de}
\affiliation{Integrated Quantum Optics, Applied Physics, University of Paderborn, 33098 Paderborn, Germany}

\author{J. Sperling}
\affiliation{Clarendon Laboratory, University of Oxford, Parks Road, Oxford OX1 3PU, United Kingdom}

\author{C. Silberhorn}
\affiliation{Integrated Quantum Optics, Applied Physics, University of Paderborn, 33098 Paderborn, Germany}

\author{W. Vogel}
\affiliation{Arbeitsgruppe Theoretische Quantenoptik, Institut f\"ur Physik, Universit\"at Rostock, D-18051 Rostock, Germany}

\begin{abstract}
	Atmospheric channels are a promising candidate to establish secure quantum communication on a global scale.
	However, due to their turbulent nature, it is crucial to understand the impact of the atmosphere on the quantum properties of light and examine it experimentally.
	In this paper, we introduce a method to probe atmospheric free-space links with quantum light on a laboratory scale.
	In contrast to previous works, our method models arbitrary intensity losses caused by turbulence to emulate general atmospheric conditions.
	This allows us to characterize turbulent quantum channels in a well-controlled manner.
	To implement this technique, we perform a series of measurements with different constant attenuations and simulate the fluctuating losses by combining the obtained data.
	We directly test the proposed method with an on-chip source of nonclassical light and a time-bin-multiplexed detection system.
	With the obtained data, we characterize the nonclassicality of the generated states for different atmospheric noise models and analyze a post-selection protocol.
	This general technique in atmospheric quantum optics allows for studying turbulent quantum channels and predicting their properties for future applications.
\end{abstract}

\date{\today}

\maketitle

\section{Introduction}

	Driven by the goal of establishing quantum key distribution in global communication networks, the field of atmospheric quantum optics has turned into a rapidly developing and growing research area over the last years.
	Starting from the successful implementations of ground-to-ground atmospheric links \cite{Ursin, Scheidl, Fedrizzi2009, Capraro, Yin, Ma, Peuntinger}, small-scale experiments \cite{Bourgoin13, Nauerth, Wang, Bourgoin15} further demonstrated the feasibility of ground-satellite communication with quantum light.
	Only recently, the first experiments with an actual satellite node have been reported \cite{Vallone15,Dequal16,Vallone16,Carrasco-Casado,Guenthner}.

	Due to temporal and spatial variations of the optical properties of the atmosphere, caused by atmospheric turbulent flows, the transmittance of such links itself varies in a turbulent manner.
	In order to understand the behavior of nonclassical light propagating through atmospheric links, it is important to perceive a profound understanding of these channels.
	A first quantum description of fluctuating atmospheric losses has been introduced in Ref. \cite{Semenov2009}.
	On this basis, advanced fluctuation loss models have been studied \cite{beamwandering,VSV2016}, which have been proven to accurately describe experiments \cite{Usenko,VSV2016}. 

	Besides a proper understanding of fluctuating free-space links, another vital aspect is to gain a profound knowledge on the action of such channels on the quantum states of light and its nonclassical properties as they offer the resource for the transmission of quantum information.
	In particular, this implies to answer the following questions. 
	{\it Which quantum states and nonclassical properties are robust against atmospheric fluctuations?} 
	{\it Under which conditions can the quantum effects be preserved?}
	Answering these questions will provide the basis for a successful implementation, optimization, and development of quantum information applications in atmosphere links, such as quantum key distribution.

	However, the uncontrollable character of turbulence itself prevents one from answering these questions in a simple way.
	Specific scenarios and effects have been addressed in theory, such as entangled Gaussian states \cite{GaussSatellites}, particular non-Gaussian states \cite{Bohmann15,NonGaussSatellites}, the influence on the photon statistics \cite{PPTD73,Milonni}, or violation of Bell inequalities \cite{Semenov2010,Gumberidze}.
	A step towards a more general treatment has been taken in Ref. \cite{BSSV16}, where entanglement conditions for the important class of two-mode Gaussian states in fluctuating loss channels have been derived.
	This approach has been further extended to test for general nonclassical effects that also apply to multimode, non-Gaussian states \cite{BSSV16b}.

	An alternative to theoretical investigations of quantum states in free-space applications is given by performing laboratory experiments in which fluctuating losses are artificially introduced.
	For this purpose, the atmospheric channels need to be properly emulated in laboratory environments in order to create realistic free-space losses.
	Possible implementations that have been exploited are turbulence cells consisting of air guns \cite{Pors,Pereira} or simulations of an atmospheric link by varying a lens position \cite{Bourgoin15b}.
	Alternatively, one can use phase screens \cite{Rodenburg,Zhang}.
	The latter introduces random phase noise which can mimic the influence of weak turbulence and affects the spatial mode structure of the propagating light field.
	Such disturbances are of great importance for applications using the orbital angular momentum degree of freedom of light; see, e.g., \cite{Arnaut,Mair}.
	Note that in those scenarios, the atmospheric noise can be transformed into loss \cite{Farias,Ndagano}.
	This phase noise influences the spatial mode structure, for example, different orbital angular momentum modes, and may lead to losses in a particular spatial mode due to scattering into other modes.
	Here, we focus on amplitude fluctuations, due to the atmospheric effects of beam wandering, beam shape deformation, beam-broadening, etc., which are the predominant effects of atmospheric losses in most experimental setups \cite{Ursin, Scheidl, Fedrizzi2009, Capraro, Yin, Ma, Peuntinger, Bourgoin13, Nauerth, Wang, Bourgoin15,Vallone15,Dequal16,Vallone16,Carrasco-Casado} and which also partially originate from light scattering.
	Yet, they can be completely described as random losses in a single optical mode propagating in a lossy, linear medium \cite{Semenov2009}.
	This also applies to scenarios which do not employ the spatial degrees of freedom of the propagating light field.
	The corresponding theory of these atmospheric fluctuating losses has been introduced in Refs. \cite{Semenov2009,beamwandering,VSV2016}.
	It is also worth mentioning that atmospheric phase fluctuations and wave front distortions can be circumvented via polarization multiplexing of a local oscillator beam and homodyne measurement; see, e.g., Ref. \cite{Peuntinger}. 
	Therefore, we focus on fluctuating intensity losses and demonstrate how they can be simulated in a controlled manner.

	In order to reliably explore such atmospheric quantum channels on a laboratory scale, one needs to have control over the random losses. 
	Therefore, it is necessary to apply the fluctuations of the transmittance in a well-defined fashion, which allows one to implement the probability distribution of the atmospheric transmittance, for example, the models in Refs. \cite{Semenov2009, beamwandering, VSV2016}.
	Afterwards, important aspects can be addressed, such as finding optimal quantum states, evaluating detection methods, identifying limitations in the performance, estimating the achievable quantum bit rate, and others.

	In this paper, we develop theoretical tools and experimentally demonstrate the simulation of arbitrary free-space channels with intensity fluctuations.
	We use measurements at different fixed attenuation levels which allows us to use a post-processing approach to weight and merge the obtained data to analyze the sought-after atmospheric model.
	This is possible as the combined data corresponds to a density operator which is estimated by an ensemble average.
	As such, it is irrelevant if the data sets were indeed taken with a fluctuating loss or measured in an ordered fashion and merged afterwards. 
	As our source of nonclassical light, we utilize an on-chip-integrated down-conversion source \cite{kruse_dual-path_2015}, which is advantageous for satellite based quantum communication because of its small size and low weight.
	We realize measurements for $100$ different attenuation levels and detect the quantum light with a time-multiplexed detector (TMD) system.
	This employs highly efficient superconducting nano-wire detectors. 
	From the recorded click-statistics, we infer the reference nonclassicality for the individual attenuation levels. 
	Then, we merge the different data sets according to different atmospheric models and directly infer the nonclassicality from the obtained click-counting statistics.
	This post-processing approach allows us to introduce the fluctuating losses in a controlled manner without the need of introducing actual experimental noise which might also affect the experiment in an undesired way. 

	The paper is structured as follows.
	In Sec. \ref{sec:discretization}, we discuss fluctuating atmospheric channel losses and introduce a discretization procedure for the distribution of the transmittance.
	Our experimental setup is described in Sec. \ref{sec:experiment}.
	In Sec. \ref{sec:click}, we study nonclassicality criteria for click-counting data and the influence of constant loss on the measured light.
	In Sec. \ref{sec:application}, applications of our channel simulations are analyzed.
	Finally, we summarize and conclude in Sec. \ref{sec:conclusion}.

\section{Simulating atmospheric losses}\label{sec:discretization}

	In this section, we establish the procedure with which fluctuating free-space losses are simulated by measurements with fixed transmittance.
	Therefore, we firstly recall the model of the fluctuating free-space channels.
	Secondly, we introduce the discretization of the corresponding probability distribution of the transmittance and show how it can be applied to measured data.

\subsection{Models for fluctuating loss channels}

	The atmosphere acts as a turbulent propagation medium for radiation fields.
	Especially, the refractive index of air changes randomly due to turbulent flows that cause temporal and spatial variations of the atmospheric properties, such as temperature and pressure.
	Therefore, the transmitted light signal is degraded by effects like beam broadening, wandering, deformation, and others.
	Besides these random variations in the atmosphere, one additionally has to account for loss caused by the finite aperture size at a receiver stage.
	Combining these two effects results in a fluctuating loss channel.

	A quantum description of such fluctuating atmospheric channels has been introduced in Ref. \cite{Semenov2009}.
	Because of fluctuating losses, the transmittance of the channel is not constant, but follows a probability distribution of the transmittance (PDT), which completely characterizes the free-space communication link.
	The action of atmospheric channels on a quantum state can be described in terms of the Glauber-Sudarshan $P$ function \cite{Glauber,Sudarshan}.
	That is, the phase-space representation of the input state, $P_\mathrm{in}$, is mapped onto the output function, $P_{\rm out}$, which is described by the integral transformation
	\begin{align}\label{eq:Pout}
		P_{\rm out}(\alpha)=\int\limits_0^1 d \eta\,\mathcal{P}(\eta)\frac{1}{\eta}P_{\rm in}\Bigl(\frac{\alpha}{\sqrt{\eta}}\Bigr),
	\end{align}
	where $\mathcal{P}(\eta)$ is the PDT and $\eta$ is the intensity transmittance.
	For different atmospheric conditions, different PDTs have been derived in Ref. \cite{beamwandering} and Ref. \cite{VSV2016} for weak and strong turbulence, respectively.
	It is important to notice that those loss models agree very well with experimental data \cite{Usenko,VSV2016}.

\subsection{Simulating atmospheric channels via discretization}

	As shown above, each atmospheric channel can be fully characterized by its PDT.
	The PDT is in general a continuous probability density, defined on the domain $[0,1]$ for the transmissivities $\eta$.
	However, in order to obtain the PDT from experiments, the transmittance is measured with a certain binning \cite{Peuntinger,Usenko} or it can be recovered from discrete single-photon click events \cite{Capraro}.
	Therefore, in many experimental scenarios one rather deals with a discretized version of PDT than with the continuous one.

	This fact motivates us to use separate measurements with fixed transmittance and a subsequent weighting of data in order to obtain a fluctuating-loss channel.
	Formally, this approach corresponds to a discretization of the PDT $\mathcal{P}(\eta)$.
	The probability mass function $\tilde{\mathcal{P}}(Y)$ of the discrete random variable $Y$ can be given as
	\begin{align}\label{eq:discretization}
		 \tilde{\mathcal{P}}(Y=\eta_k)=\mathcal{P}(\eta_k)\left[\sum\limits_{j=0}^{n}\mathcal{P}(\eta_j)\right]^{-1}.
	\end{align}
	This means that the PDT is evaluated on a sufficiently dense and finite subset of equidistant supporting points $\eta_k= k/n$ (for $k=0,\ldots,n$) in the interval $[0,1]$.
	For details on this type of discretization and other possibilities, we refer to the review in Ref. \cite{discretization}.
	Note that we will label all discrete probability mass functions with a tilde ($\tilde{\mathcal P}$) in the further course of this paper.

	If one uses such a discretization, it is important to assess its quality, i.e., to quantify how well the discretized version is consistent with the original distribution.
	One way of evaluating the quality of the discretization is by comparing the moments (moments of the transmittance) of the discretized distribution with the original one.
	We will apply this particular approach in this paper.

	With the discretized PTD, we can weight the separate constant-loss measurements according to $\tilde{\mathcal{P}}(Y)$.
	We denote the different measurement sets by $x_j$, where $j$ indicates the transmittance $\eta_j$.
	Then, we obtain the data set including the simulated atmospheric losses by
	\begin{align}\label{eq:atmosData}
		x^{\rm atm.}=\sum\limits_{j=0}^{n}\tilde{\mathcal{P}}(\eta_j)x_j.
	\end{align}
	In order to guarantee the validity of this approach, each individual measurement run has to be comparable in the sense that all parameters are equal (e.g., same optical devices, equal measurement time, or equal number of repetitions) except for the transmittance of the channel.

	Note that such an approach is possible as the obtained data correspond to a density operator, which is experimentally estimated by an ensemble average.
	In the case of an atmospheric channel, this includes the average over different attenuations.
	Therefore, it is irrelevant if either the data sets are indeed taken in a fluctuating fashion or are constructed from a sufficiently large set of measurements with fixed attenuations.
	In particular, in the case of pulsed light, each individual pulse of the experiment suffers from a fixed attenuation and the fluctuating character arises only from the ensemble average.  
	Hence, the measurement of pulses itself is already an averaging of discrete measurement outcomes of the form in Eq.~\eqref{eq:atmosData}.

\section{Experimental setup}\label{sec:experiment}

\begin{figure}[ht]
	\includegraphics[width=.8\columnwidth]{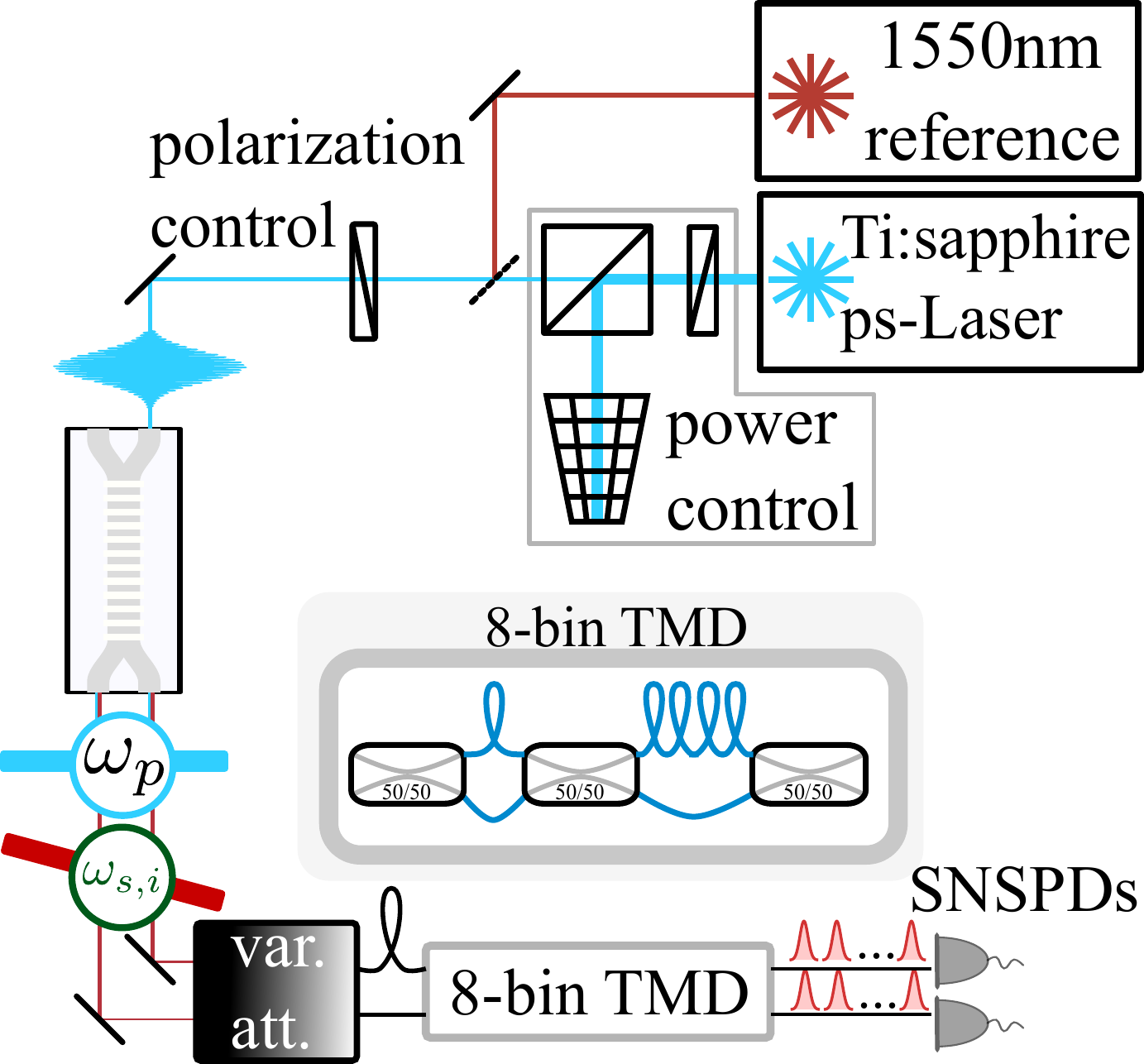}
	\caption{(Color online)
		Experimental setup.
		We employ the dual-path down-conversion source~\cite{kruse_dual-path_2015} as a nonclassical light source. 
		To simulate the effect of atmospheric noise, the light undergoes 100 different attenuation levels, realized with a combination of half-wave plate and polarizing beam splitter. 
		Finally, the states impinge on an eight-bin time-multiplexed detector (TMD) and are recorded by superconducting nano-wire single-photon detectors (SNSPDs).
		As a classical reference, we use attenuated laser pulses from a $1550$\,nm laser diode.
	}\label{fig:setup}
\end{figure}

	The experimental implementation is outlined in Fig. \ref{fig:setup}.
	To simulate the effect of atmospheric fluctuating losses on nonclassical states, we use a dual-path down-conversion source which has been characterized in Ref. \cite{kruse_dual-path_2015}. 
	The source emits two identical uncorrelated single-mode squeezed light fields with a broad frequency spectrum
	\begin{align}
		&|\zeta\rangle\otimes|e^{i\pi}\zeta\rangle=\left[\hat S\otimes\hat S^\dag\right]|0\rangle\otimes|0\rangle,
		\text{ with }\\\nonumber
		&\hat S{=}\exp\!\left[-\frac{i\mathcal{B}}{\hbar}\int_{\omega_\mathrm{min}}^{\omega_\mathrm{max}}d\omega d\omega^{'}
		\left(f(\omega,\omega^{'}) \hat{a}^\dagger(\omega) \hat{a}^\dagger(\omega^{'}){+}\mathrm{h.c.} \right)\right]
	\end{align}
	where $\int_{\omega_\mathrm{min}}^{\omega_\mathrm{max}} d\omega d\omega^{'}$ denotes the integration over the frequency band that impinges on the detectors and $\hat{a}^\dagger(\omega)\otimes \hat 1$ and $\hat 1\otimes \hat a^\dagger(\omega)$ represent the creation operators in waveguide 1 and 2, respectively, of the dual-path source at frequency $\omega$. 
	We describe the strength of the squeezing operation in terms of the optical gain $\mathcal{B}$. 
	In terms of photon number statistics $(p_n)$ this means that the state in one mode fulfills $p_n=0$ for odd $n$ and $p_n>0$ for even $n$.
	Although we record both spatial modes with the TMD, we restrict ourselves to one of the modes for the simulation of atmospheric turbulences, since the states in both spatial modes are identical and show the same behavior.
	Yet, let us mention that our approach is not restricted to single-mode states of light.

	It is also worth mentioning that our integrated source has tremendous advantages for satellite communication.
	It is small, light, relatively cheap, and inherently stable due to the waveguide geometry.
	Additionally, it can be easily integrated into a fiber-based network.
	
	To operate our source, we pump it with picosecond laser pulses from a Ti:sapphire laser at $\lambda_p\approx758$\,nm that undergo a power and polarization control. 
	After the source, the generated photon pairs are cleaned from the pump and unwanted background in the telecom regime.
	To emulate the noise effects of the atmosphere, we implement our previously proposed method and measure the transmitted quantum state for $n=100$ different attenuation levels $\eta$ in both arms of the source.
	We achieve this via a variable half-wave plate and a polarizing beam splitter.
	At full transmission of this attenuation stage, we find a detection efficiency of the overall setup to be approximately $22\%$.
	After the attenuation, the quantum states are detected with a detection system that consists of an eight-bin ($N=8$) time-multiplexing network \cite{FJPF03, ASSBW03, RHHPH03, ASSBWFJPF04} and two superconducting nano-wire detectors.
	We choose the power of the pump field such that we record on average $2.7$ clicks per pulse at a full transmission of the attenuation stage, $\eta=1$.
	Thus, we are operating in the few-photon regime, which is of particular interest for free-space experiments; see, e.g., Refs.~\cite{Ursin, Scheidl, Fedrizzi2009, Capraro, Yin, Ma,Vallone15,Dequal16,Vallone16}.

	Moreover, to compare the effects of noise on nonclassical states with a classical reference, we couple a pulsed diode laser at $1550$\,nm (32$\,$ps pulse duration, $100$\,kHz repetition rate) through the setup.
	Analogously to the quantum case, we record the attenuated classical photon statistics of the produced coherent state for $100$ different attenuation levels.

\section{Nonclassicality criteria and constant loss measurements}\label{sec:click}
	
	In this section, we explain the used nonclassicality criteria and apply them to the measured click-counting data.
	First, we recall how one can test for nonclassicality directly from the click-counting statistics.
	Second, we will investigate the behavior of the nonclassical light recorded for the case of constant attenuations.
	
\subsection{Click-counting nonclassicality criteria}

	In order to determine the nonclassical character of the light propagating through an atmospheric channel, we use criteria based on the measured click-counting statistics \cite{SVA12,SVA12a,SVA13}. 
	Note that detection schemes at the single-photon level recently have been demonstrated to be applicable for free-space links even at bright daylight \cite{Liao16}.
	In our experiment, we measure the light with a TMD with $N=8$ time bins from which we directly obtain the click statistics.
	Such a system is described by a click-counting probability $c_k$, with $k$ indicating the number of clicks ($0\leq k\leq N$). 
	The normalized click-counting statistics resulting from such a measurement follows the quantum version of a binomial distribution \cite{SVA12,SVA12a,SVA13},
	\begin{align}\label{eq:click_statistics}
		c_{k}=&\langle{:} \binom{N}{k}\hat m^{N-k}(\hat 1-\hat m)^{k}{:}\rangle,
	\end{align}
	where ${:}\,\cdot\,{:}$ indicates the normal-ordering prescription.
	The normally ordered expectation value of the operator $\hat m=\exp\{-\hat \Gamma(\hat n)\}$ yields the no-click probabilities, where $\hat \Gamma$ and $\hat n$ denote the detector-response function and the photon-number operator, respectively.
	
	We directly use the recorded click-counting statistics in order to identify quantum features of the detected light.
	It has been shown that the matrix of moments $M^{(K)}$ is non-negative for any classical light field \cite{SVA13},
	\begin{align}\label{Eq:ClMoM}
		0\leq M^{(K)}{=}(\langle{:}\hat m^{s{+}t}{:}\rangle )_{(s,t)},
	\end{align}
	with $s,t=0,\ldots, K/2\leq N/2$ for even $K$ and $N$.
	The superscript $(K)$ defines the highest moment within the matrix $M$.
	Its simplest form is given by moments up to the second order
	\begin{equation}\label{eq:secondorder}
		M^{(2)}=\begin{pmatrix} 1 & \langle{:}\hat m{:}\rangle \\ 
		\langle{:}\hat m{:}\rangle & \langle{:}\hat m^2{:}\rangle \end{pmatrix}.
	\end{equation}
	If the matrix $M^{(K)}$ is not positive semidefinite, the detected light field is nonclassical. 
	As we can directly sample the required moments from the measured data \cite{SVA13}, we do not have to perform an additional data post-processing which is the inherent strength of the method.
	
\subsection{Constant losses}\label{sec:constant}
	
	We start our data analysis of the measured click data with fixed attenuations. 
	This analysis will serve as a reference for the eventual case of simulated fluctuating loss.
	In order to test for nonclassical effects with the matrix of moments \eqref{Eq:ClMoM}, one needs to choose which moments one wants to consider.
	We have already mentioned the simplest possible matrix of moments in Eq. \eqref{eq:secondorder} for $K=2$.
	Yet we can use moments up to the eighth order, defining the $5\times 5$ matrix $M^{(8)}$ in Eq. \eqref{Eq:ClMoM}, because we use an $N=8$ bin TMD.
	In order to identify that the matrices are not positive semidefinite, i.e., they show nonclassical features, it is sufficient to show that they have at least one negative eigenvalue \cite{SBVHBAS15}. Hence, we consider the minimal eigenvalues $e^{(2)}$ and $e^{(8)}$ of the matrices of moments $M^{(2)}$ and $M^{(8)}$, respectively.
	We stress that for classical light, the constraint \eqref{Eq:ClMoM} applies, which is identical to $e^{(K)}\geq0$.
	
\begin{figure}[ht]
	\includegraphics[width=0.85\columnwidth]{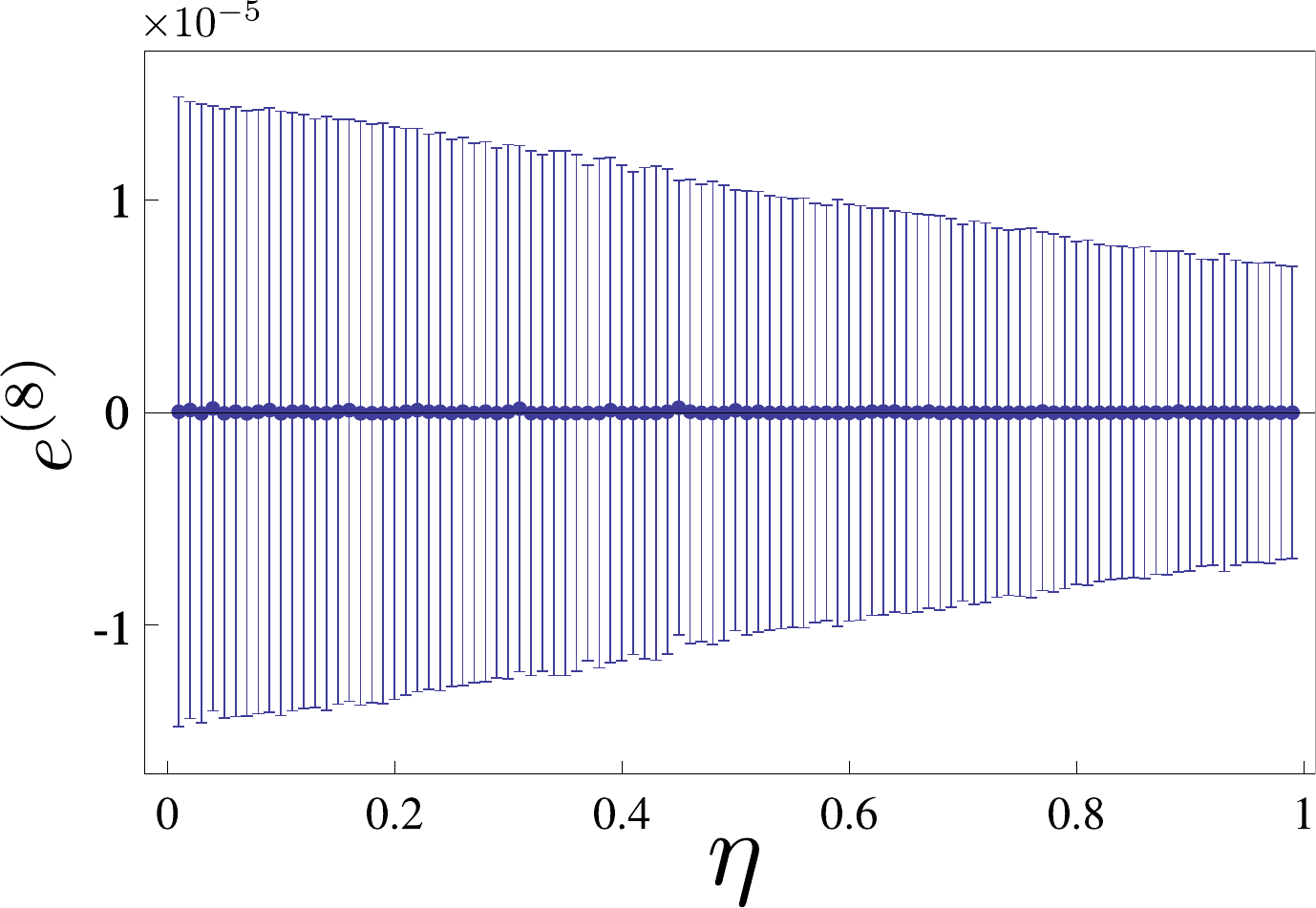}
	\caption{(Color online)
		The minimal eigenvalue $e^{(8)}$ of the classical coherent-state reference for different attenuation levels $\eta$. 
		As the eigenvalue is consistent with zero for all measurements, our detection scheme works as expected.
		Due to the fact that the no-click events increase with decreasing $\eta$, the errors of the eigenvalue also increase.
	}\label{fig:classical}
\end{figure}

	In order to ensure that our setup does not introduce fake nonclassical effects, we first consider the classical coherent-state reference signal.
	The used reference signal has an intensity of less than one photon per pulse.
	For all realized attenuations, the minimal eigenvalues $e^{(8)}$ of the classical signal are consistent with zero within the error bar as it can be observed in Fig. \ref{fig:classical}.
	This is the expected result for a coherent-state reference \cite{SVA13}, and we can conclude that our detection scheme works appropriately and does not herald false nonclassicality.
	Note that a similar measurement with varying intensities has been recently applied to accurately calibrate the used detection device \cite{BKSSV16}, which additionally determines the quantum efficiency.

\begin{figure}[t]
	\includegraphics[width=0.85\columnwidth]{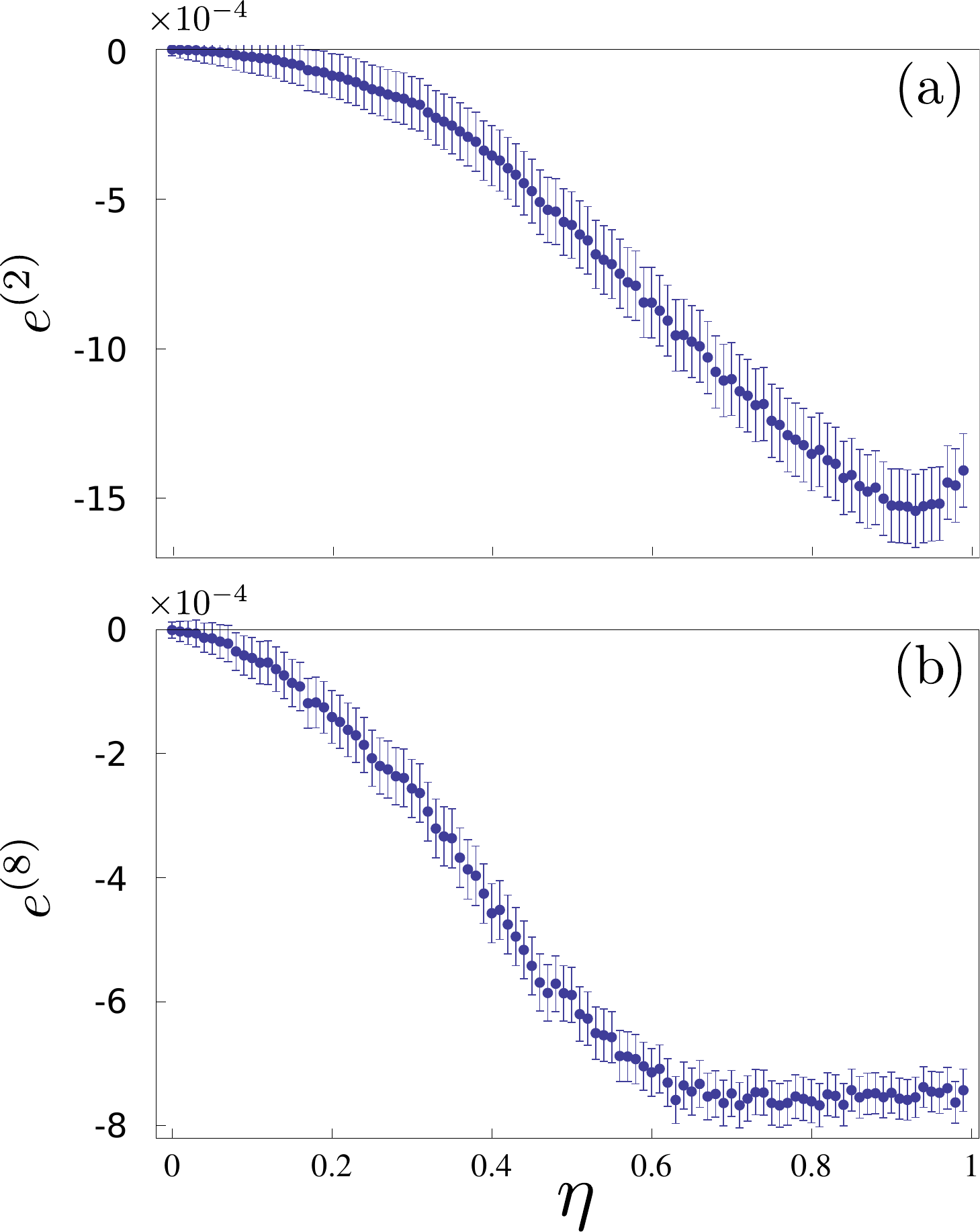}
	\caption{(Color online)
		Minimal eigenvalues $e^{(K)}$ for the second-order [panel (a), $K=2$] and eighth order [panel (b), $K=8$] matrix of moments. 
		In both cases, we observe nonclassicality with a high significance for a large range of attenuation parameters $\eta>0$.
	}\label{fig:ConstantLoss}
\end{figure}

	Let us now consider the results for the non-classical light from the on-chip source. 
	In Fig. \ref{fig:ConstantLoss}, we plot the minimal eigenvalues and their errors bars depending on the constant attenuation $\eta$. 
	The upper plot corresponds to the second-order matrix $M^{(2)}$ from Eq. \eqref{eq:secondorder} and the lower plot shows the results for the eighth-order matrix $M^{(8)}$. 
	For both, we directly obtain the confirmation that our source emits nonclassical light, as both show clear negativities for $\eta\approx1$.
	Let us emphasize that even without the additional attenuation the overall efficiency is given by the detection efficiency of $22\%$.

	Furthermore, we obtain negative eigenvalues for all measured transmissivities for the nonclassical light in Fig. \ref{fig:ConstantLoss}.
	However, not for all values of attenuation can we verify the nonclassicality with high significance when considering the error margin.
	For $e^{(2)}$ and $e^{(8)}$ we can determine the nonclassical character with at least three standard deviations for $\eta\gtrsim 0.34$ and $0.18$, respectively. 
	In this sense, the matrix $M^{(8)}$ and its corresponding minimal eigenvalue $e^{(8)}$ are more sensitive to the quantum features. 
	This by itself is an interesting result as it demonstrates that by taking into account higher-order moments, the sensitivity of nonclassicality detection can be perceptibly improved.
	Note, a similar behavior can be found for any other $e^{(K)}$ with $K<N=8$.
	Therefore, we will solely apply the matrix $M^{(8)}$ and its minimal eigenvalue $e^{(8)}$ in the further course of this paper.

\section{Application: Simulation of different fluctuating loss channels}\label{sec:application}

	Using the tools introduced so far, we can study any fluctuating-loss scenario described by a PDT.
	In the scope of this paper, we restrict ourselves to the simulation of two realistic atmospheric channels. 
	In addition, to underline that our approach works for arbitrary fluctuating loss scenarios, we apply the method to a family of two-parameter probability mass functions at the end of this section.
	This shows that our method can treat any fluctuating-loss scenario, especially new theoretical models or experimental conditions can be directly tested in this way.
	Hence, the versatility and general applicability of the method are demonstrated.
	
\subsection{Strong turbulence: Log-normal model}

	As the first example, we investigate the influence of strong atmospheric turbulence, which can be approximately described by the log-normal distribution.
	However, it overestimates the high-transmission behavior, which is more correctly described in Ref.~\cite{VSV2016}.
	The log-normal model was used, e.g., in Refs. \cite{Capraro,PPTD73,Milonni}.
	The corresponding PDT reads
	\begin{equation}\label{eq:LN}
		\mathcal{P}_{\rm LN}(\eta)=\frac{1}{\eta \sigma \sqrt{2\pi}}\exp{\frac{-(\ln\eta- \mu)^2}{2\sigma^2}},
	\end{equation}
	which is characterized by the location parameter $\mu\in\mathbb{R}$ and scale parameter $\sigma>0$, which determine the moments of the distribution $\langle \eta^s \rangle=\exp (s\mu+s^2\sigma^2/2)$.
	In the following, we consider the specific case $\mu=-1.75$ and $\sigma^2=0.55$ \cite{VSV2016}.
	
\begin{figure}[ht]
		\includegraphics[width=0.85\columnwidth]{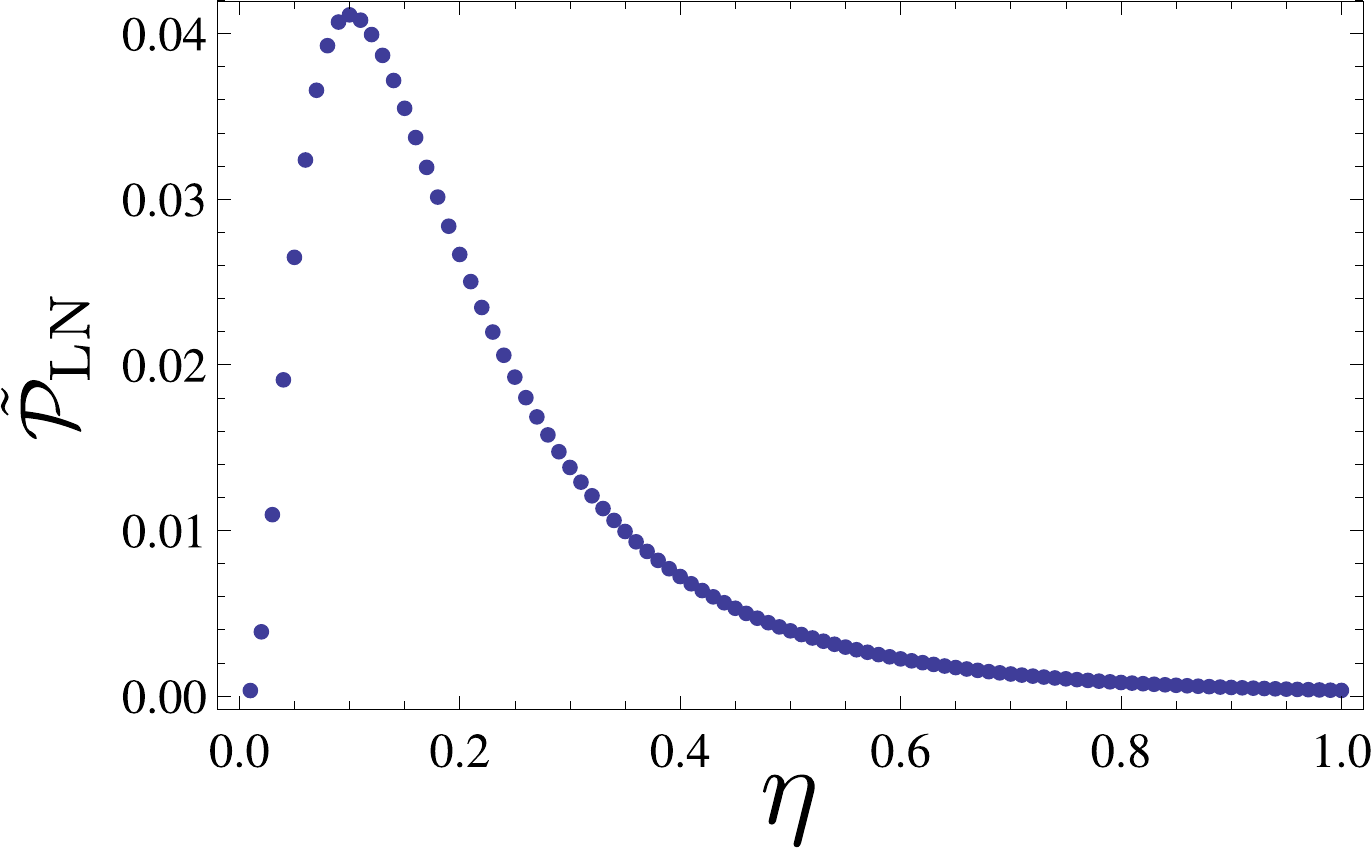}
		\caption{(Color online)
			Discretization of the log-normal distribution \eqref{eq:LN}.
		}\label{fig:disLN}
	\end{figure}

	From Eq. \eqref{eq:LN}, we get a discretized distribution $\tilde{\mathcal P}$ according to Eq. \eqref{eq:discretization}.
	The resulting probability mass function $\tilde{\mathcal{P}}_{\rm LN}(\eta)$ is displayed in Fig. \ref{fig:disLN} for our $100$ supporting points.
	In order to assess the quality of the discretization, we compare the moments of $\tilde{\mathcal{P}}_{\rm LN}(\eta)$ with the moments of the original $\mathcal{P}_{\rm LN}(\eta)$.
	For the first three moments of the transmittance, i.e., its mean, variance, and skewness, we find relative errors of $0.6\times10^{-3}$, $2.1\times10^{-3}$, and $4.7\times10^{-3}$, respectively.
	This demonstrates the high quality of the discretization and justifies the effort to collect and analyze the data for the large number of different attenuations.
	According to this distribution, we create the resulting atmospheric click statistics using Eq. \eqref{eq:atmosData}.

	The data evaluation for our generated nonclassical state is performed first.
	In particular, we can immediately test if the nonclassical character can be still observed for the turbulence model under study.
	Therefore, we use the violation of the classicality condition~\eqref{Eq:ClMoM} where the moments $\hat m^l$ are replaced with the atmospheric sampling formula
	\begin{equation}
		\hat m^l_{\rm atm}=\sum_{j=0}^{n}\tilde{\mathcal{P}}(\eta_j)\sum_{k=0}^{N-l}\frac{\binom{N-k}{l}}{\binom{N}{l}}c_k(\eta_j),
	\end{equation}
	with $c_k(\eta_j)$ [see Eq.~\eqref{eq:click_statistics} and Ref. \cite{SVA13}] being the click statistics recorded under the attenuation $\eta_j$.
	According to this replacement, we use the matrix $M^{(8)}_{\rm atm}$, and we calculate its minimal eigenvalue and find $e^{(8)}_{\rm atm}=(2.34\pm 0.05)\times 10^{-2}$, which is clearly positive.
	Hence, we cannot identify the nonclassicality anymore.

	This is an unexpected result as we obtain negative values for $e^{(8)}_{\rm atm}$ for all fixed attenuations; see Fig. \ref{fig:ConstantLoss}.
	Therefore, one might suspect that a weighted combination of the constant losses [see Eq. \eqref{eq:atmosData}] would also yield a negative eigenvalue $e^{(8)}_{\rm atm}<0$.
	The fact that we cannot observe nonclassicality highlights that such atmospheric losses show a non-trivial behavior and need to be investigated with care under controlled conditions before one performs expensive experiments in open-air test sites.
	Particularly, fluctuating loss channels cannot simply be approximated by a constant loss channel with the same mean transmittance.
	Otherwise, we could indeed verify nonclassicality for this case.
	In fact, the action of the fluctuating loss yields a statistical mixture of the input state [see Eq.~\eqref{eq:Pout}], which might have lost all the nonclassical features of the original nonclassical input state.

\begin{figure}[ht]
	\includegraphics[width=0.85\columnwidth]{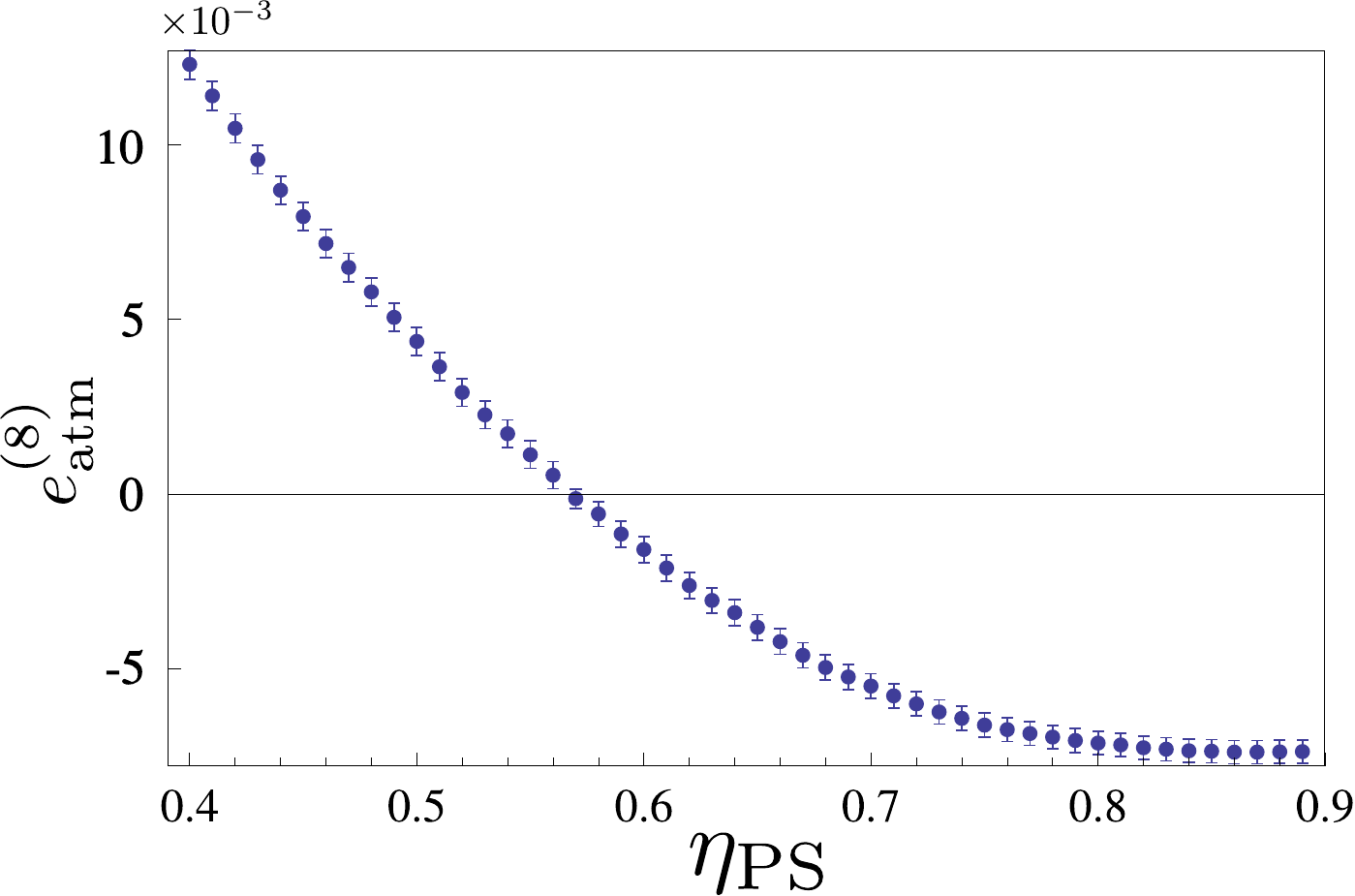}
	\caption{(Color online)
		The minimal eigenvalue of the simulated atmospheric matrix of moments $e^{(8)}_{\rm atm}$ is plotted as a function of the post-selection cutoff transmittance $\eta_{\rm PS}$.
		The minimal eigenvalue decreases for increasing $\eta_{\rm PS}$. 
		For values $\eta_\mathrm{PS}>0.59$, we verify the nonclassical behavior of the detected light.
	}\label{fig:PS}
\end{figure}

	As we do not obtain nonclassical behavior of the detected light with atmospheric noise, we need to find a way to retain the possibility for secure communication with such a quantum resource. 
	One possibility to achieve this is to use the post-selection protocol \cite{Peuntinger} which has been proposed in Ref.~\cite{beamwandering}. 
	This protocol consists of a monitoring of the atmosphere and considering only events for which the transmittance is above a certain threshold value $\eta_\mathrm{PS}$. 
	The results for the application of this post-selection protocol are depicted in Fig. \ref{fig:PS}. 
	We plot the minimal eigenvalues $e^{(8)}_{\rm atm}$ for different post-selection threshold values $\eta_\mathrm{PS}$. 
	For $\eta_\mathrm{PS}\gtrsim 0.59$, we regain the nonclassical character with significances of more than three standard deviations. 
	This shows the versatility of our simulation approach not only for the simulation of atmospheric channel data, but also for the determination of possible post-processing protocol parameters.

\subsection{Beam wandering: Log-negative Weibull distribution}

	As a second example, we study the influences of atmospheric beam wandering on the quantum features of the light from our nonclassical light source.
	Beam wandering is the dominating effect in weakly turbulent channels \cite{beamwandering}.
	Due to the atmospheric turbulence, the beam is wandering across the receiver aperture plane causing fluctuations in the transmittance.
	The corresponding PDT for a single, spatial mode is consistent with a log-negative Weibull distribution \cite{beamwandering,Usenko}.

	In the following, we analyze the transfer of the non-classical light depending on the turbulence strength of such a channel. 
	The strength of the turbulence is determined by the Rytov parameter $\sigma_\mathrm{R}^{2}$. 
	A value $\sigma_\mathrm{R}^{2}=0$ corresponds to a non-turbulent scenario, weak turbulence is characterized by $0<\sigma_\mathrm{R}^{2}<1$, and for $\sigma_\mathrm{R}^{2}\approx 1\dots 10$, we obtain moderate turbulence. 
	For our analysis, we use the same parameters as for a $1.6$-km-long channel \cite{VSV2016}.
	As discussed before, we discretize the PDT in order to apply it to our measured data.
	The relative errors of this discretization in the first three moments of the transmittance are $0.75\%$, $1.48\%$, and $2.20\%$.

\begin{figure}[ht]
	\includegraphics[width=.9\columnwidth]{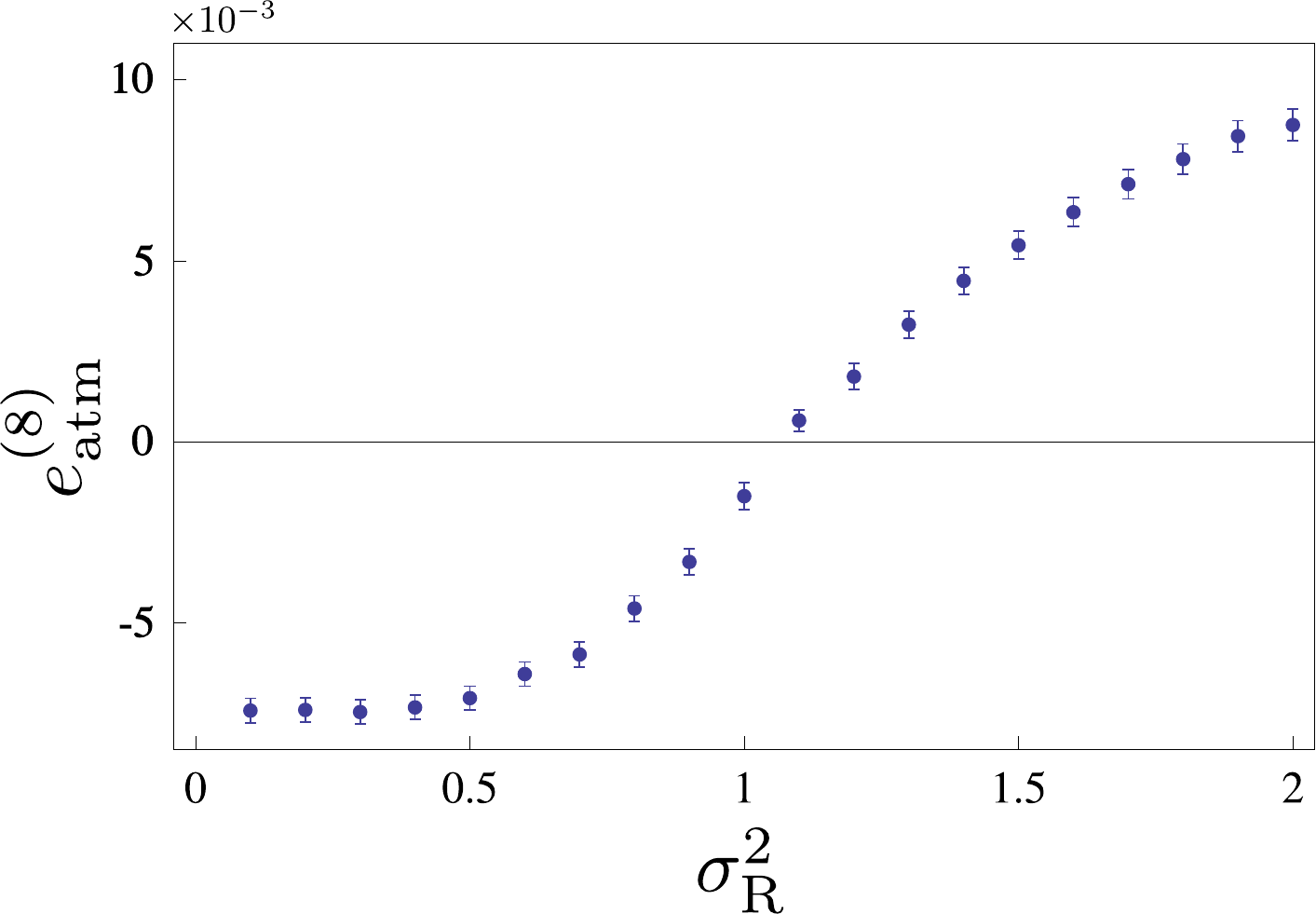}
	\caption{(Color online)
		The plot shows the dependence of the nonclassicality test, $e^{(8)}_{\rm atm}<0$, on the turbulence strength determined by the value of the Rytov parameter $\sigma_{\rm R}^{2}$.
		Nonclassicality is significantly preserved for $\sigma_{\rm R}^{2}\lesssim 1$.
	}\label{fig:BW}
\end{figure}

	We weight the measured data according to the beam wandering PDT for different strengths of the Rytov parameter $\sigma_\mathrm{R}^2$. 
	Figure \ref{fig:BW} summarizes the result of this analysis. 
	We obtained the minimal eigenvalues $e^{(8)}_{\rm atm}$ from the click statistics which is subjected to different strengths of beam wandering.
	We find that the nonclassicality can be preserved up to a turbulence strength of $\sigma_\mathrm{R}^{2}\approx 1$. 
	Hence, the nonclassicality of our generated states vanishes for moderate turbulence strengths. 
	With this type of analysis, we have shown that we can simulate the influence of atmospheric turbulence conditions for a full-scale atmospheric transmission setup in the laboratory and retain the tolerable strength of turbulence for a given quantum state in free-space links.

\subsection{General test: Beta-binomial distributions}

	Finally, we explore how our method performs for PDTs following arbitrary probability functions.
	As a general example, we consider a family of probability distributions, the beta-binomial distributions, which are in our case defined as
	\begin{equation}\label{eq:betadis}
		\tilde{\mathcal{P}}_{\rm BB}(\eta_k|n,\alpha,\beta)=\binom{n}{\eta_k n}\frac{{\rm B}(\eta_k n+\alpha,n-\eta_k n+\beta)}{{\rm B}(\alpha,\beta)},
	\end{equation}
	where ${\rm B}(a,b)=\Gamma(a)\Gamma(b)/\Gamma(a+b)$ is the beta distribution [$\Gamma(x)$ is the $\Gamma$ function] and $\eta_k=k/n$ denotes the value of discrete transmittance with $k{=}0,\dots, n$.
	The beta-binomial distribution is a binomial distribution in which the success probability for each trial follows the beta distribution in a random fashion.
	We also emphasize that the distribution $\tilde{\mathcal{P}}_{\rm BB}(\eta|n,\alpha,\beta)$ is already discrete and is determined by the three positive parameters $n$, $\alpha$, and $\beta$. 
	In our case, we have $n=100$, since we measured $100$ separate attenuation levels and vary the two parameters $\alpha$ and $\beta$ to cover a whole family of different probability distributions.
	For the distribution \eqref{eq:betadis}, the mean is $n\alpha/(\alpha+\beta)$ and the variance reads $n\alpha\beta(\alpha+\beta+n)/[(\alpha+\beta)^2(\alpha+\beta+1)]$, which allows one to uniquely relate the parameters $\alpha$ and $\beta$ to the first and second moments of the transmissivities.
	
\begin{figure}[ht]
	\includegraphics[width=.85\columnwidth]{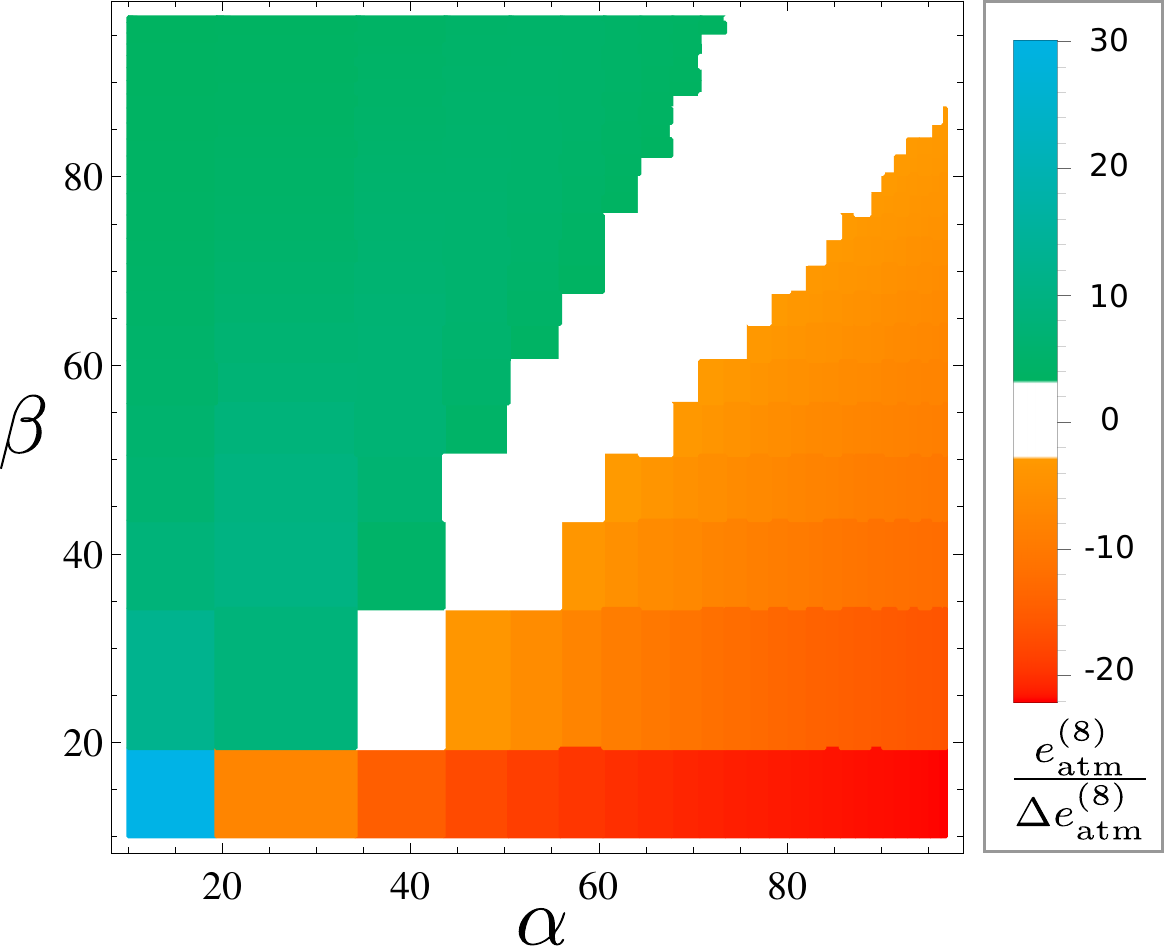}
	\caption{(Color online)
		Nonclassicality analysis for the general family of beta-binomial distributions. 
		We plot the signed significance of the minimal eigenvalue, $e^{(8)}_{\rm atm}/\Delta e^{(8)}_{\rm atm}$, as a function of the parameters $\alpha$ and $\beta$.
		White areas indicate insignificant results, whereas a green-blue and orange-red tone represent classical and nonclassical features, respectively.
	}\label{fig:beta}
\end{figure}

	For our analysis, we weight the data according to $\tilde{\mathcal{P}}_{\rm BB}(\eta_k|n,\alpha,\beta)$ with different $\alpha$ and $\beta$ and analyze the minimal eigenvalues $e^{(8)}_{\rm atm}$.
	The results for this family of distributions are given in Fig. \ref{fig:beta}. 
	We use the signed significance of the minimal eigenvalues, i.e., the ratio between the minimal eigenvalue and its error bar, $e^{(8)}_{\rm atm}/\Delta e^{(8)}_{\rm atm}$, as a figure of merit.
	This quantity is non-negative for classical states.
	For a wide range of parameters, we identify regions for which we can certify nonclassicality (orange-red) for the channel described by $\tilde{\mathcal{P}}_\mathrm{BB}(\eta_k|n,\alpha,\beta)$.
	There are also regions for which the nonclassicality is fully degraded (green-blue) by the channel.

	The analysis of such a general probability mass function demonstrates the general applicability of our approach.
	It also shows that our method is not restricted to already known PDTs.
	In fact, we can test in this way any loss scenario which is described by a PDT that might be derived in the future and which covers various forms of atmospheric conditions.

\section{Conclusion}\label{sec:conclusion}

	We introduced and implemented a method to simulate general fluctuating loss channels for modeling various atmospheric conditions in well-controlled laboratory experiments.
	We use measurements with discrete, constant losses and a subsequent weighting and merging of the data according to different atmospheric loss channels. 
	We formulated a mathematical framework for this approach and evaluated relative errors in the discretization routine.

	We applied our approach to measured data of non-classical light from an on-chip integrated source.
	The resulting quantum light was detected by a state-of-the-art click counting device.
	The nontrivial impact of the atmosphere on the nonclassicality was studied.
	Further, we could demonstrate the practicability of this procedure for various atmospheric loss conditions covering the ranges of weak and strong turbulence.
	This included the analysis of a post-selection protocol. 
	Beyond that, we also directly demonstrated the general applicability by simulation of a whole family of probability distributions.

	We want to emphasize that the proposed laboratory-based simulation of atmospheric channels is not limited to particular detection schemes.
	It can be adapted to other kind of measurements, such as homodyne detection.
	Furthermore, the accuracy can be improved by increasing the number of measurements with constant transmission coefficients, which corresponds to a finer sampling of the probability distribution of the atmospheric transmittance.
	Finally, one could also include other noise effects, such as dephasing, in order to account for other possible disturbances.
	Again, the latter can be achieved via constant modifications of the phase and a subsequent mixing of the resulting data according to a given dephasing model.
	Thus, we believe that our approach is an versatile tool to assist the development of atmospheric quantum optics and, in particular, to predict the success of quantum key distribution via free-space links.

\section*{Acknowledgements}
	The authors thank A. A. Semenov and D. Yu. Vasylyev for valuable discussions.
	The project leading to this application has received funding from the European Union’s Horizon 2020 research and innovation programme under Grant No. 665148.
	M. B. and W. V. are grateful for financial support by Deutsche Forschungsgemeinschaft through Project No. VO 501/22-1.

\end{document}